 \documentclass{article}

\usepackage[dvips]{graphicx}
\usepackage{amsmath}
\usepackage{amsfonts}
\usepackage{amssymb}
\usepackage[latin1]{inputenc}
\usepackage{xcolor}

\title{Hanbury Brown and Twiss, Hong Ou and Mandel effects and other landmarks in quantum optics: from photons to atoms}

\author{Alain Aspect \thanks{Institut d'Optique Graduate School, Universit\'e Paris Saclay, 91120 Palaiseau}}
\date{}

\begin{document}

\maketitle

\noindent Published as Chapter 12  in the proceedings of the Les Houches Summer School 1996 : {\it Current Trends in Atomic Physics.} Edited by Antoine Browaeys, Trey Porto, Charles S. Adams, Matthias Weidemuller, and Leticia F. Cugliandolo. Oxford University Press (2019). 

\noindent DOI: 10.1093/oso/9780198837190.003.0012


%
%
%
%
%
%
%

\tableofcontents

%
\newpage
For a long time, the name ``quantum optics'' referred  mostly to the physics of lasers. In fact, in a laser, what is fully quantum is the amplifying medium itself, with the quantized levels of the emitter (atoms, molecules, ions), but light can be perfectly described as a classical electromagnetic wave, as it is done in the Lamb theory of the laser\cite{lamb1964theory}. In fact, in 1960, there was only one phenomenon involving visible light that demanded the use of the quantum theory of light: it was spontaneous emission. Absorption or stimulated emission could  be described by the semi-classical classical theory of matter-light interaction [see for instance \cite{grynberg2010introduction}], in which matter only is quantized. And when it came to  freely propagating light, the description as classical electromagnetic waves was found perfectly adequate, provided that one used the semiclassical model of photo detection, and that a statistical description were used to describe incoherent light, such as thermal radiation, or light emitted by discharge lamps.

It is only in the 1960's that emerged the idea that it might  be important to describe freely propagating light as a quantized  system, involving in particular the notion of photon. When Roy Glauber took the challenge of using the quantum formalism to describe the Hanbury Brown and Twiss (HBT) effect, he had to develop a  formalism, which not only allowed him to describe the HBT effect, but also  was available, from then on, to allow physicists to render an account of new genuine quantum optics effects that were discovered in the next decades.  Among these effects, whose description and understanding demands quantization of freely propagating electromagnetic field,
one must  cite the ones involving pairs of entangled of photons, which were used to show a violation of Bell's inequalities [see references in \cite{Aspect:1999gm,aspect2015viewpoint}], and  the Hong Ou and Mandel  (HOM) effect\cite{Hong:1987co}. There were other quantum effects, related to properties of single photons\cite{aspect:2017mooc1photon}, such as photon anti-bunching  in resonance fluorescence\cite{Kimble:1977kk}, or  anticorrelation for a single photon on a beam-splitter\cite{Grangier:1986ie}. But in this lecture I will  put the emphasis on the HBT and the HOM effects, which have been recently revisited in our laboratory with atoms replacing photons. These effects are remarkable landmarks in quantum optics since their description demands to use  the notion of  two photon amplitudes interference, which is a major ingredient\footnote{The other ingredient of the second quantum revolution is the experimental ability to observe and manipulate individual quantum objects, and the Quantum Monte-Carlo methods that suggest clear intuitive images for the evolution of these individual quantum objects. See \cite{dowling2003quantum,Aspect:2004introductionsuqm} . } of the second quantum revolution\cite{dowling2003quantum,Aspect:2004introductionsuqm}. 

Today, I will first present my views on the second vs the first quantum revolution, then describe the Hanbury Brown and Twiss effect with photons, and indicate why it was so important in the development of modern quantum optics. The presentation of our experiments on the HBT effect with atoms will allow me to emphasize the analogies but also the increased richness of the effect when going from photons to atoms. I will similarly describe the HOM effect for photons and its significance, and then present the analogous experiment with atoms. In conclusion, I will put these two effects in the long list of landmarks in the development of quantum optics, and indicate what has been done and what remains to be done with atoms in lieu of photons.
\section{Two great quantum mysteries} \label{s1}
In the early 1960's, in chapter 1 of volume 3 of his famous lectures on physics\cite{Feynman:1963kf}, Feynman described wave particle duality as ``the only quantum mystery ''. Two decades later, however,  in a paper that is considered the founding paper of quantum information\cite{Feynman:1982yx}, he recognized that there was another great mystery, entanglement. Why are these two extraordinary features of quantum mechanics different in nature? 

Wave particle duality refers to a single quantum particle, which can be described both as a wave and a particle. Each of these descriptions involves a classical notion: a wave propagating in the ordinary space-time, or a particle whose trajectory is developed in the ordinary space-time. What is  quantum, and hard to swallow, is that these two  descriptions are used for the same object, which belongs a priori to one of the two categories: an electron, a neutron, is a priori a particle, but we must also think of it as a wave; light is a priori a wave, but we must also think of it as composed of particles, the photons. But each of these  behaviors can be described  without any problem, in the usual ordinary space-time. 

In contrast, entanglement between several particles must be described in an abstract Hilbert space, which is the tensor product of  the spaces of each of the entangled objects. A problem may arise, then, when one wants to  give an image of  what happens in our ordinary space-time. For instance, for two maximally entangled particles separated in space,  violation of Bell's inequalities is predicted by quantum mechanics and observed experimentally\cite{Aspect:1999gm,aspect2015viewpoint}. In that situation,  any image in our ordinary space-time involves either negative probabilities, or some-degree of non-locality, i.e., a contradiction with the notion that nothing can propagate faster than light. Both sides of the alternative are very hard to swallow, as stressed by Feynman\cite{Feynman:1982yx}: ``\textit{I've entertained myself always by squeezing the difficulty of quantum mechanics into a smaller and smaller place, so as to get more and more
worried about this particular item. It seems to be almost ridiculous that you
can squeeze it to a numerical question that one thing is bigger than another.
But there you are--it is bigger than any logical argument can produce, if
you have this kind of logic.}''
 It could be tempting to content oneself with the observation that there is no non-locality  in the Hilbert space where the two entangled particles are described. But as emphasized by Asher Peres, a famous quantum optics theorist: ``\textit{Quantum phenomena do not occur in a Hilbert space. They occur in a laboratory}''[page 373 in\cite{peres1995quantum}].

Because it is difficult to renounce locality or positiveness of probabilities, phenomena based on entanglement are much more difficult to swallow than the ones based on wave particle duality. This is why, in this lecture,  I will focus on two quantum optics landmarks that we have recently revisited with atoms, in which entanglement must be invoked to give a consistent quantum description.
\section{The Hanbury Brown and Twiss effect for photons}\label{s2}  
The experiment reported by  R Hanbury Brown and RQ Twiss in 1956\cite{brown1956correlation}  is  considered the landmark signaling the beginning of modern quantum optics. It is indeed for giving a fully consistent description of that experiment that R Glauber developed the Quantum Optics formalism that we still use today.\footnote{According to Claude Cohen-Tannoudji (private communication) the interpretation of the Forrester et al. experiment\cite{forrester1955photoelectric}  on quantum beats in light emitted by a spectral lamp had provoked intense discussions, which had prepared the minds to the necessity of a fully consistent quantum description of such phenomena. }

\subsection{\label{ss21} Experimental observation} 
\begin{figure}[h!] 
\centering
\includegraphics[width=0.8\linewidth]{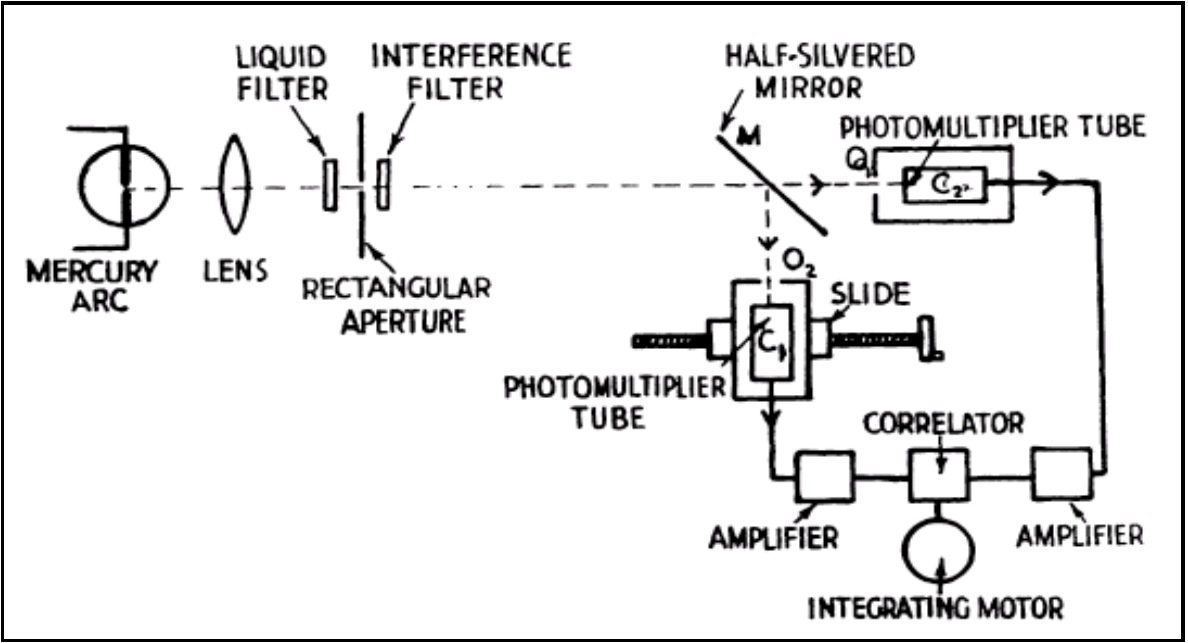}
\caption{Schematic view of the original HBT experiment.}
\label{Figure_HBT}
\end{figure}

Figure \ref{Figure_HBT} describes the original experiment, which was a study of the  intensity fluctuations of light emitted by an incoherent source (laser had not yet been invented). Two photomultipliers, almost image of each other by reflection in a beam splitter, allowed one to monitor the correlation function of the photocurrents associated with light detection at $(\mathbf{r}_1, t)$ and $(\mathbf{r}_2, t+\tau)$, with $(\mathbf{r}_1, t)$ and $(\mathbf{r}_2, t+\tau)$ as close of each other than one wants. One also monitors the average photocurrent at each detector.

According to the semi-classical theory of quantum optics, the photocurrent is proportional to the light intensity, i.e., the squared modulus of the classical complex electric field
\begin{equation}
i({\bf{r}},t) \propto I({\bf{r}},t) = \left |E^{(+)}({{\bf{r}}},t)\right|^2 = E^{(-)}({{\bf{r}}},t)\,E^{(+)}({{\bf{r}}},t) \,,
\label{eq1}
\end{equation}
so that the normalized current correlation function is equal to the normalized correlation function of the light intensity:

\begin{equation}
{g^{(2)}}({{\bf{r}}_1},{{\bf{r}}_2};\tau ) = \frac{{\left\langle {i({{\bf{r}}_1},t)\,i({{\bf{r}}_2},t + \tau )} \right\rangle }}{{\left\langle {i({{\bf{r}}_1},t)} \right\rangle \left\langle {i({{\bf{r}}_2},t)} \right\rangle }}= \frac{{\left\langle {I({{\bf{r}}_1},t)\,I({{\bf{r}}_2},t + \tau )} \right\rangle }}{{\left\langle {I({{\bf{r}}_1},t)} \right\rangle \left\langle {I({{\bf{r}}_2},t)} \right\rangle }} \,.
\label{eq2}
\end{equation}
Note that in spite of a somewhat misleading, but traditional, notation,  $g^{(2)}$ is in fact a fourth order correlation function of the classical complex  electric field
\begin{equation}
{g^{(2)}}({{\bf{r}}_1},{{\bf{r}}_2};\tau ) = \frac{\left\langle E^{(-)}({{\bf{r}_1}},t)\,E^{(+)}({{\bf{r}_1}},t) E^{(-)}({{\bf{r}_2}},t+\tau)\,E^{(+)}({{\bf{r}_2}},t+\tau)  \right\rangle}
{\left\langle E^{(-)}({{\bf{r}_1}},t)\,E^{(+)}({{\bf{r}_1}},t) \right\rangle
\left\langle E^{(-)}({{\bf{r}_2}},t+\tau)\,E^{(+)}({{\bf{r}_2}},t+\tau)  \right\rangle} \,.
\label{eq2_N1}
\end{equation}

\begin{figure}[h!] 
\centering
\includegraphics[width=0.8\linewidth]{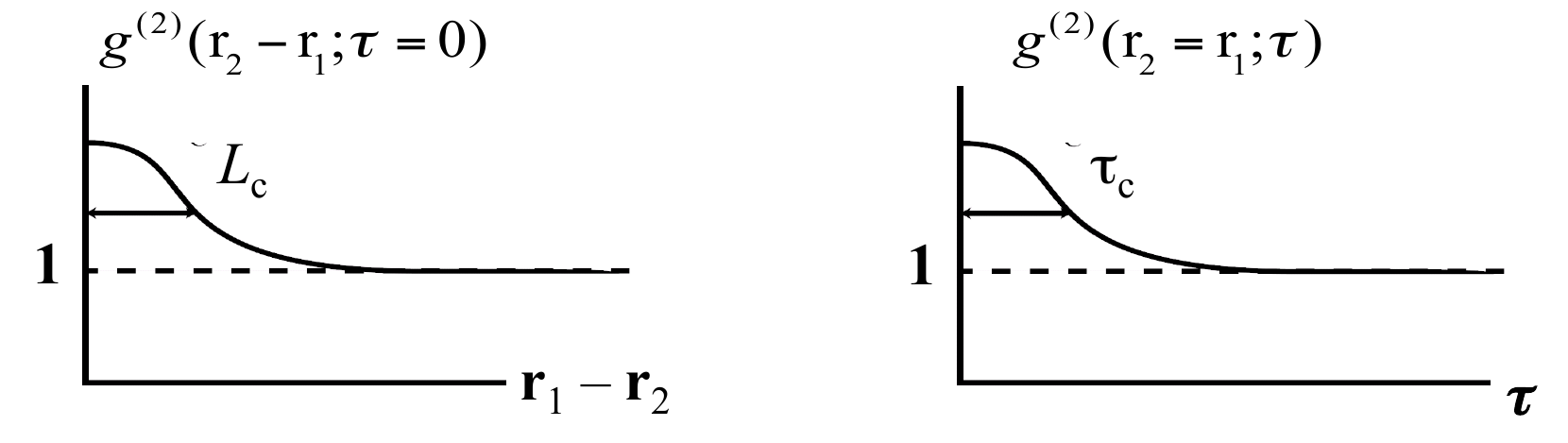}
\caption{Results of the original HBT experiment. The normalized correlation function is maximum, with a value of 2, at zero distance and delay, where it characterizes intensity fluctuations. It drops to the value of 1, which means no correlation, for a delay larger than the correlation time $\tau_c$, or a distance larger than the correlation length $L_\mathrm{c}$.}
\label{Figure_HBTresults}
\end{figure}

Figure \ref{Figure_HBTresults} shows the results of the original HBT experiment. At zero distance and time delay, the normalized correlation function is nothing else than the average of the square of the intensity. The value greater than 1 for $\mathbf r_1=\mathbf r_2 $ and $\tau = 0$ means that  light intensity fluctuates. More precisely, the value of 2 indicates that the variance $ \langle I^2 \rangle - \langle I \rangle^2$  is equal to the squared average intensity $\langle I \rangle^2$. At long distance ($|\mathbf r_1 - \mathbf r_2| \gg L_\mathbf c $) and/or large delay ($\tau \gg \tau_\mathbf c$), the autocorrelation function drops to 1, which means no correlation between the fluctuations.

\begin{figure}[h!] 
\centering
\includegraphics[width=0.7\linewidth]{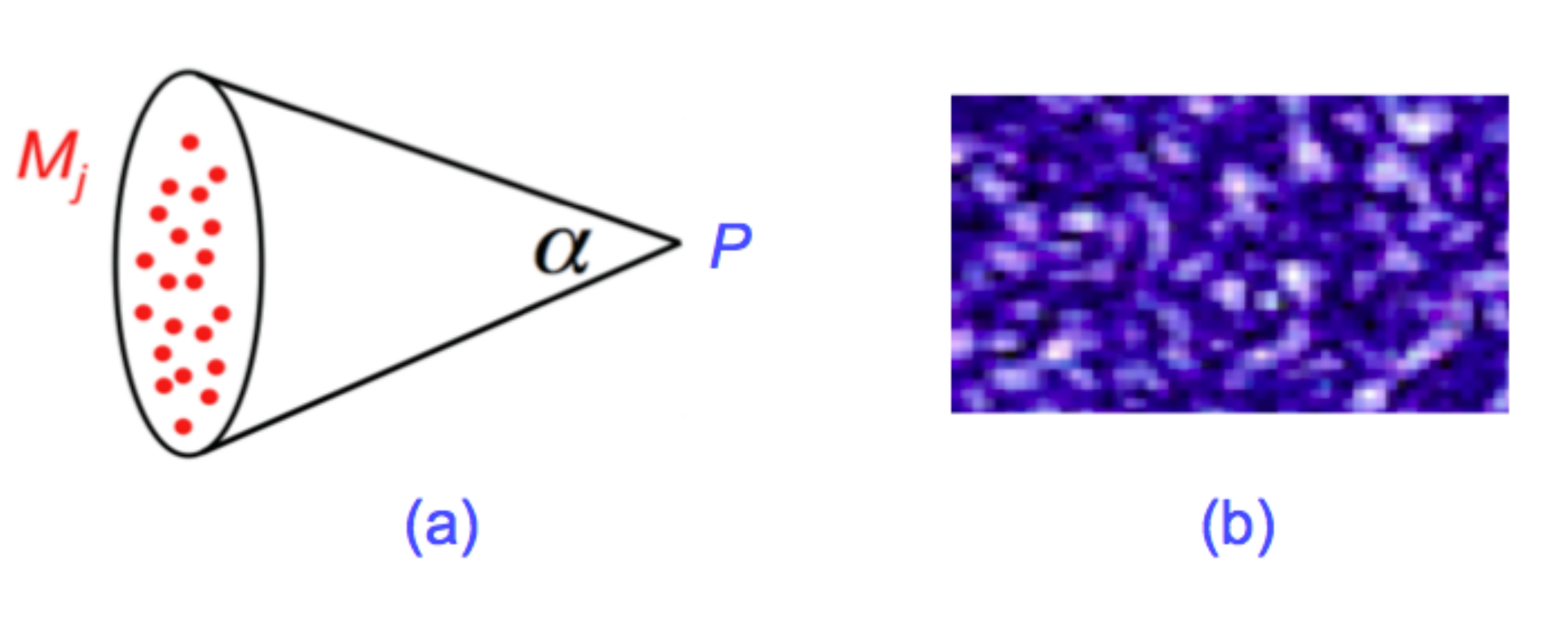}
\caption{Intensity pattern produced by an incoherent source: (a) The source, composed of many independent emitters, is seen from the detection point under an angular diameter $\alpha$. (b) At a given time, the intensity pattern is a speckle pattern, whose ``grains'' have a characteristic size of the order of $L_\mathrm{c} =\lambda / \alpha$;  this pattern is a random process, which evolves with a characteristic time $\tau_\mathrm{c}$.}
\label{Figure_sourceHBT}
\end{figure}

The goal of HBT was to perform such a measurement on the light emitted by a star, in order to determine its angular diameter $\alpha$. The reason is that the correlation length $L_\mathrm{c}$ is linked to the angular diameter under which one sees the star (Figure \ref{Figure_sourceHBT}) by the relation
\begin{equation}
L_\mathrm{c} = \frac{\lambda}{\alpha}
\label{eq3}
\end{equation}
where $\lambda$ is the wavelength of the light. This would allow them to measure angular diameters of star, a measurement made impossible with standard astronomical methods by the atmospheric fluctuations.

\subsection{\label{ss22} Semi-classical interpretation} 
We describe classically the field emitted by the source as the sum of many contributions issued from independent emitters $j$ in the source. The complex electric field at $P$ (Figure \ref{Figure_sourceHBT})  is thus 
\begin{equation}
E^{(+)}(P,t) = \sum\limits_j {{a_j}\exp \left\{ {{\phi _j} + \frac{{{\omega _j}}}{c}{M_j}P - {\omega _j}t} \right\}} 
\label{eq4}
\end{equation}
where the $\phi _j$ are independent random variables. The field $E^{(+)}(P,t)$ is a sum of many random variables with the same statistical properties. It is thus a Gaussian random process, as a consequence of the Central Limit Theorem. We can then use the Gaussian Moment Theorem   to express ${g^{(2)}}({{\bf{r}}_1},{{\bf{r}}_2};\tau )$, which is a fourth order moment of the complex electric field (Equations \ref{eq2_N1}), as 
\begin{equation}
{g^{(2)}}({{\bf{r}}_1},{{\bf{r}}_2};\tau ) = 1 + {\left| {{g^{(1)}}({{\bf{r}}_1},{{\bf{r}}_2};\tau )} \right|^2}
\label{eq5}
\end{equation}
where 
\begin{equation}
g^{(1)}({{\bf{r}}_1},{{\bf{r}}_2};\tau ) =\frac{\left\langle E^{(-)}({{\bf{r}}_1},t)E^{(+)}({{\bf{r}}_2},t+\tau)\right\rangle}{\left\langle E^{(-)}({{\bf{r}}_1},t)E^{(+)}({{\bf{r}}_1},t)\right\rangle^{1/2}\left\langle E^{(-)}({{\bf{r}}_2},t+\tau)E^{(+)}({{\bf{r}}_2},t+\tau)\right\rangle^{1/2}}
\label{eq6}
\end{equation}
is the second order moment of the complex electric field. In fact, $g^{(1)}$ is the so-called first order coherence function, whose spatial and temporal widths are respectively the coherence length $L_{\mathrm c}$ and the coherence time $\tau_{c}$. Within a factor of the order of 1, the functions $g^{(1)}({{\bf{r}}_1},{{\bf{r}}_2};\tau )$ and $g^{(2)}({{\bf{r}}_1},{{\bf{r}}_2};\tau )$ have thus the same widths.
Since $g^{(1)}({{\bf{r}}_1}-{{\bf{r}}_2} = 0 \, ; \tau=0 ) =1$, one has $g^{(2)}({{\bf{r}}_1}-{{\bf{r}}_2} = 0 \, ; \tau=0 ) =2$. This factor of 2 is characteristic of a Gaussian process. 

Note in passing an illuminating interpretation of the widths of $g^{(2)}({{\bf{r}}_1},{{\bf{r}}_2};\tau )$. Let us think of the intensity pattern produced around $P$ by the source of Figure \ref{Figure_sourceHBT}. At any given time, it is a speckle pattern, whose ``grains'' have a characteristic size of the order of $L_\mathrm{c} =\lambda / \alpha$;  this pattern is a random process, which evolves with a characteristic time $\tau_\mathrm{c}$. If two detectors are separated by less than the grain size, the detected intensities are correlated fluctuating quantities. For a larger separation, the fluctuations of the detected intensities are uncorrelated.

\subsection{\label{ss23} A hot debate} 
When HBT proposed to build what they called ``an intensity interferometer'' to measure $g^{(2)}$ in order to deduce stars angular diameters, their application to get funding was rejected\cite{brown1974intensity}, based on the following argument. 

\begin{figure}[h!] 
\centering
\includegraphics[width=1.0\linewidth]{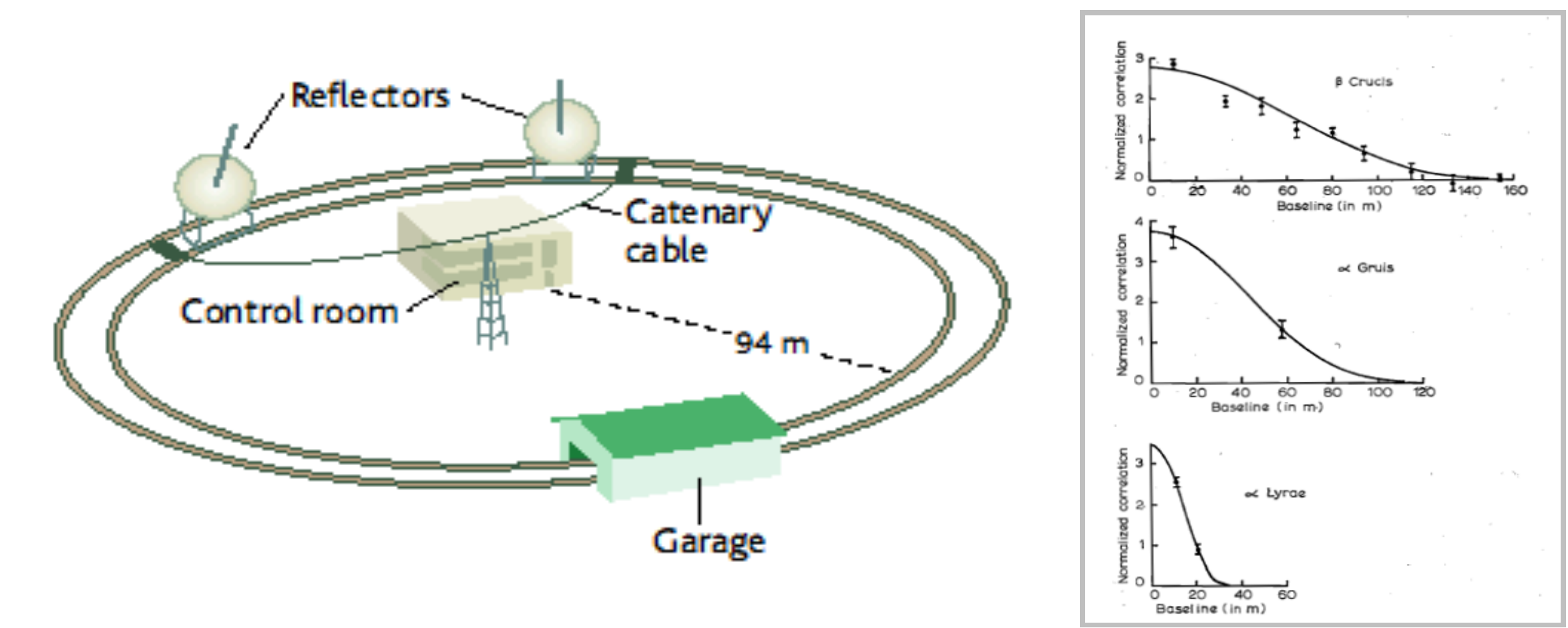}
\caption{The intensity interferometer built in Australia by R Hanbury Brown et al. The intensity correlation function could be measured up to separations of 188 m. The right panel shows some examples, which allowed them to determine the angular diameter of stars of the southern hemisphere. }
\label{Figure_HBTaustralie}
\end{figure}

Let us think of the experiment with the two detectors working in the photon counting mode\cite{Rebka:1957oy}. The correlation function ${g^{(2)}}({{\bf{r}}_1},{{\bf{r}}_2};\tau )$ is then expressed as a function of single and joint detection probabilities 
\begin{equation}
g^{(2)}({{\bf{r}}_1},{{\bf{r}}_2};\tau ) =
\frac{\pi^{(2)}({{\bf{r}}_1},t\,;{\bf{r}}_2,t+\tau)}
{ \pi^{(1)}({{\bf{r}}_1},t)\cdot\pi^{(1)}({{\bf{r}}_2},t+\tau)} \,.
\label{eq7}
\end{equation}
A value of 2 for $g^{(2)}$ would then mean that the photons ``come in pairs'', a totally inacceptable hypothesis, according to the referees, since photons emitted at different, possibly very distant, points of a star, are obviously independent. In spite of their efforts to argue with the referees, including the realization of the table top experiment of reference \cite{brown1956correlation}, they had to move to Australia, to find support, and build an observatory in the desert of Narrabri, where they measured the angular diameter of several stars of the southern hemisphere (Figure \ref{Figure_HBTaustralie}).

Beyond its interest in astronomy, the HBT experiment is now celebrated as the landmark whose quantum interpretation prompted the development of the modern quantum optics formalism, by Roy Glauber\cite{glauber1963photon,glauber1963coherent,glauber1963quantum,glauber1965houches}.

\subsection{\label{ss24} Quantum interpretation} 
\begin{figure}[h!] 
\centering
\includegraphics[width=0.9\linewidth]{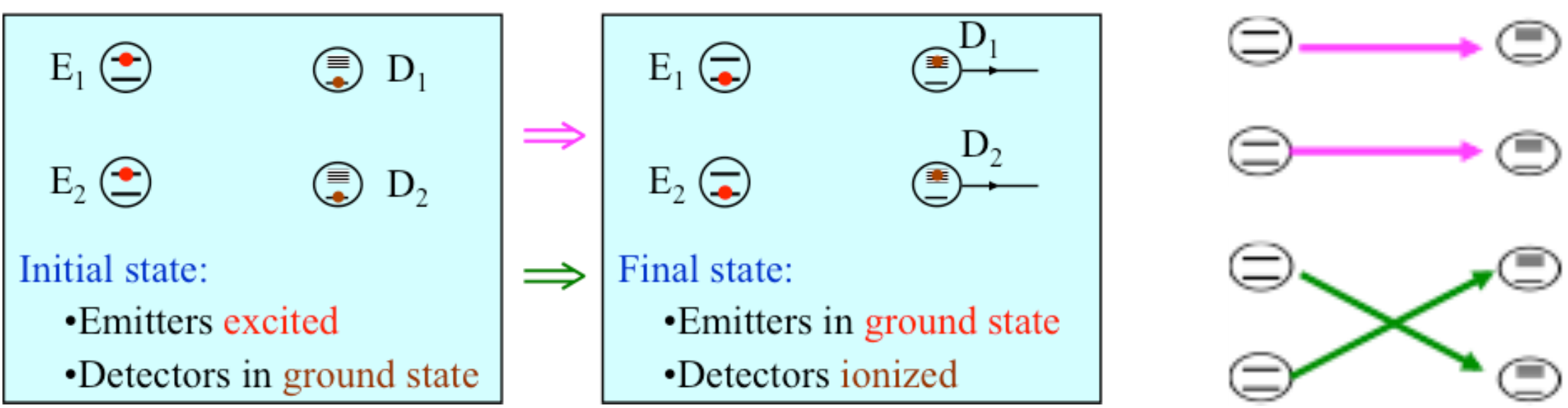}
\caption{Toy model to understand how  two photons amplitudes interference plays a role in the joint detections at  D$_1$ and D$_2$. One must add the amplitudes associated with the two processes sketched on the right panel, which correspond to the same initial and final states of the emitters and detectors. If the detectors are close enough of each other that  the two  photons wave packets are  indistinguishable, the two amplitudes have almost the same phase factor and the interference is constructive.}
\label{Figure_toymodel}
\end{figure}

From a quantum point of view,  the HBT effect is related to the quantum statistics of bosons, which tend to be detected in pairs if they cannot be distinguished. The quantum statistics is automatically taken into account in the formalism of Glauber, which is a version of second quantization well adapted to the case of photons.
I will not recall here the full treatment of reference \cite{glauber1963photon}, and will only emphasize the role of two-photon amplitudes interference, or equivalently, in that case\footnote{It must be recalled that while indistinguishability of quantum particles leads, in a first quantization point of view, to entangled states, the reciprocal is not true: entanglement can also happen between fully distinguishable particles.},  of entanglement, in the quantum description of the HBT correlations. This can be done using a toy model introduced by Glauber in his Les Houches course of 1964\cite{glauber1965houches}, and shown on Figure \ref{Figure_toymodel}.

In this model, two one-photon  wave-packets, emitted by two independent excited atoms, will overlap and be detected by two detectors. The full process consists of an evolution from an initial state to a final state, shown on Figure \ref{Figure_toymodel}. One can see by simple inspection that there are two paths to go from the initial state to the final state. These two paths are sketched on the right panel of Figure \ref{Figure_toymodel}. In order to calculate the probability of a joint detection at D$_1$ and D$_2$, one must add the amplitudes of these paths, before taking the squared modulus of the sum:
\begin{equation}
\pi^{(2)}({{\bf{r}}_1};{\bf{r}}_2)=
\left | \langle \mathrm{D}_1|U|\mathrm{E}_1\rangle  \langle \mathrm{D}_2|U|\mathrm{E}_2\rangle +  \langle \mathrm{D}_2|U|\mathrm{E}_1\rangle  \langle \mathrm{D}_1|U|\mathrm{E}_2\rangle \right |^2 \,.
\label{eq8}
\end{equation}
This is an example of a two-photon amplitudes interference effect. It is deeply linked to the notion of entanglement since the state of the photons between the emitters and the detectors is 
\begin{equation}
\left|\Psi \right\rangle= \frac{1}{\sqrt2}  \left( \left|1_{\mathrm{E}_1\mathrm{D}_1},1_{\mathrm{E}_2\mathrm{D}_2}  \right\rangle + \left|1_{\mathrm{E}_1\mathrm{D}_2},1_{\mathrm{E}_2\mathrm{D}_1}  \right\rangle \right)  
\label{eq9}
\end{equation}
where $|1_{\mathrm{E}_1\mathrm{D}_1}\rangle$ refers to one photon travelling from $\mathrm{E}_1$ to $\mathrm{D}_1$, etc... 

When one considers many different pairs of emitters, the phases  of the two terms that interfere in Equation (\ref{eq8}) differ by a random quantity, and the result is a Gaussian random process, with fluctuations such that  relation (\ref{eq5}) holds. But when the two detectors are within a coherence volume associated with the source, then the phase differences remain small compared to 1 radian, and all interferences are constructive, hence the factor of 2. 

\subsection{\label{ss25} A paradoxical situation} 
As shown in subsection \ref{ss23}, and  emphasized as early as 1956 by E. Purcell\cite{purcell1994reproduced}, the HBT effect can fully be described by a semi-clasical model in which light is not quantized. Moreover, one must admit that the quantum description is  more involved than the semi-classical one. It is thus remarkable that, in order to answer a question which could have been considered a simple curiosity rather than a necessity, R. Glauber developed a full fledged formalism, which would turn out to be necessary to interpret and analyze the genuine quantum effects that would appear later. 
\section{The Hanbury Brown and Twiss effect for atoms}\label{s3}  
\subsection{\label{ss31} From light to atoms}
In section \ref{s2}, we have seen that the interpretation of the HBT effect  demands the quantum notion of two-photons amplitudes interference, and entanglement, if light is considered as made of photons. It is therefore interesting to consider the HBT effect for other kinds of particles. As a matter of fact, as described in \cite{baym1998physics}, the HBT effect has been observed with nuclear particles, and used in order to determine collision cross sections between these particles. At the other end of the energy scale, ultra-cold atoms offer nowadays exquisite experimental methods allowing physicists to revisit the photon quantum optics experiments. Following a pioneering experiment, which demonstrated the effect using metastable Neon atoms\cite{Yasuda:1996po}, we decided to start a systematic program of study of quantum effects related to atoms entanglement, using metastable helium atoms,  the workhorse of our program of Quantum Atom Optics. 
\subsection{\label{ss32} Metastable Helium: the workhorse of Quantum Atom Optics}
\begin{figure}[h!] 
\centering
\includegraphics[width=0.8\linewidth]{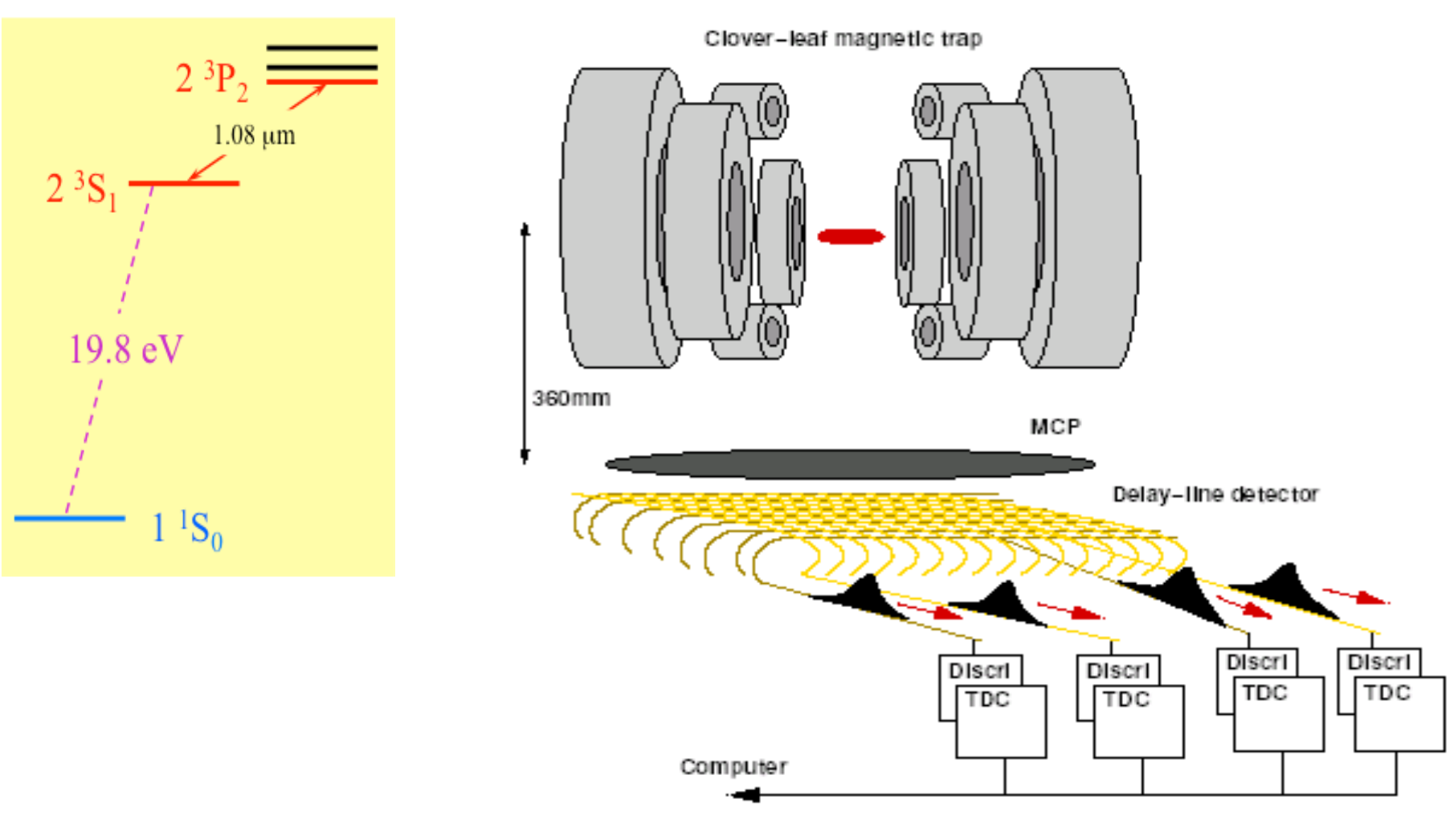}
\caption{Metastable Helium He*. Left panel: Radiative transitions from triplet levels $2\,^3P_2$ or $2\,^3S_1$ to the singlet ground state $1\,^1S_0$ are forbidden. Level $2\,^3S_1$, which is the lowest triplet state, is thus a metastable state, which plays the role of an effective ground state for atoms in any triplet state interacting with light, including  when they emit spontaneous photons. The transition at 1.08~$\mu$m can thus be used for cooling and trapping He* atoms. Right panel: When a He* atom in a triplet state falls on the Micro Channel Plate (MCP), a transition to the ground state $1\,^1S_0$ happens, and at least 19.8 eV of energy is released. This is more than enough to extract an electron from the upper face of the MCP. After multiplication in the MCP, a macroscopic electric pulse emerges on the lower face of the MCP, and is divided in four pulses propagating along delay lines, so that one can register the time and the location of the detection.   }
\label{Figure_He}
\end{figure}

As explained in the caption of Figure \ref{Figure_He}, Helium atoms in a triplet state can be manipulated with light, and thus laser cooled and trapped. 
When they are released on the MCP, they can be detected individually, with the position and time of detection of each atom recorded. Since all the free falling atoms arrive on the MCP with almost the same  velocity, time can be converted into a vertical position in the free falling cloud, and the 3D ensemble of positions in an individual cloud can be reconstructed.

The emergence of modern Quantum Optics had been permitted by the development, after World War 2, of photon counting techniques, which allowed pioneers to measure correlation functions $g^{(2)}$ in light.  MCP with He* offers similar possibilities;  better in fact, since our system with delay lines is equivalent to $10^5$ independent detectors (the MCP has a diameter of 70~mm, and the resolution is 0.2~mm), while landmark Quantum Optics experiments where performed with two detectors only. We show now how that system was used to study atomic HBT. 

\subsection{\label{ss33} Atomic HBT}
\begin{figure}[h!] 
\centering
\includegraphics[width=0.7\linewidth]{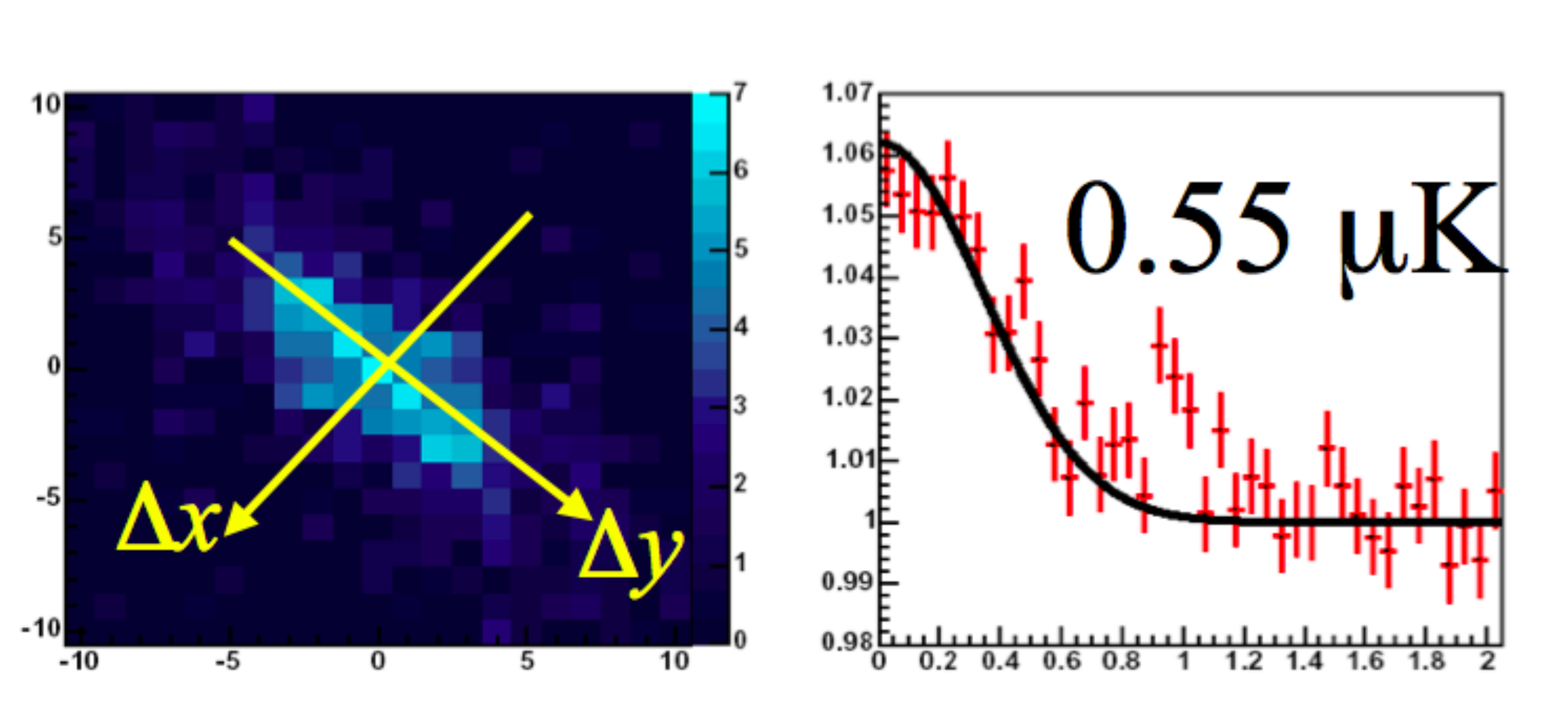}
\caption{Correlation function $g^{(2)}(\mathbf{\Delta r})$ for an initial thermal cloud at 0.55$~\mu$K, with a cigar shape elongated along $x$. The $g^{(2)}(\mathbf{\Delta r})$ function is found symmetrical by rotation around the $x$ axis, as expected, and its shape, which is shown on the cut in the $(x,y)$ plane, corresponds to the Fourier transform of the shape of the initial cloud. The maximum value of 1.06 rather than 2 is a consequence of the finite resolution of the detector, which has a point spread function wider than the atoms distribution along $x$ but narrower than the distribution along $y$. }
\label{Figure_HBTresult}
\end{figure}

Figure \ref{Figure_HBTresult} shows the result of the experiment reported in \cite{Schellekens:2005ql}, whose ingredients have been sketched in subsection \ref{ss32}.
A thermal cloud of ultracold atoms is dropped onto the detector, and we register the 3D positions of the atoms. More precisely, we define 3D pixels centered around positions $\mathbf{r}_i$ and count how many atoms we find in each pixel (actually the number is most of the time 0 and sometimes 1). We can then determine the probability to have pairs of atoms  separated by $\mathbf{\Delta r}=\mathbf{r}_i-\mathbf{r}_j$, in the whole cloud. Dividing by the product of the probabilities of having one atom in each pixel, we obtain the correlation function for one cloud
\begin{equation}
g^{(2)}_\mathrm{1cloud}(\mathbf{\Delta r}) =
\frac{ \pi^{(2)}(  \mathbf{\Delta r})}
{ [\pi^{(1)}]^2} \,.
\label{eq10}
\end{equation}
where the probabilities are defined for 1 cloud. 

The result is usually quite noisy, but  averaging $g^{(2)}_\mathrm{1cloud}(\mathbf{\Delta r})$ over many clouds yields the correlation function $g^{(2)}(\mathbf{\Delta r})$ with a good signal to noise ratio, as shown on Figure \ref{Figure_HBTresult}, which is extracted from  \cite{Schellekens:2005ql} where one can find more details.

In that reference, we also show that, when the temperature of the initial cloud is lowered below the transition temperature, one obtains a Bose Einstein Condensate (BEC), for which the correlation function is found flat, as it was the case for laser light\cite{arecchi1966time}. A similar result has been obtained with Rubidium atoms extracted from a BEC \cite{Ottl:2005ai}, while \cite{hodgman2011direct} reports on measurements of third order correlation functions in a BEC of He* atoms.

Beyond these proofs of principle, the atomic HBT effect can be used as a tool to probe many-body states of ultracold atoms\cite{Altman:2004ac}. 

\subsection{\label{ss34} Fermionic HBT effect}
\begin{figure}[h!] 
\centering
\includegraphics[width=0.5\linewidth]{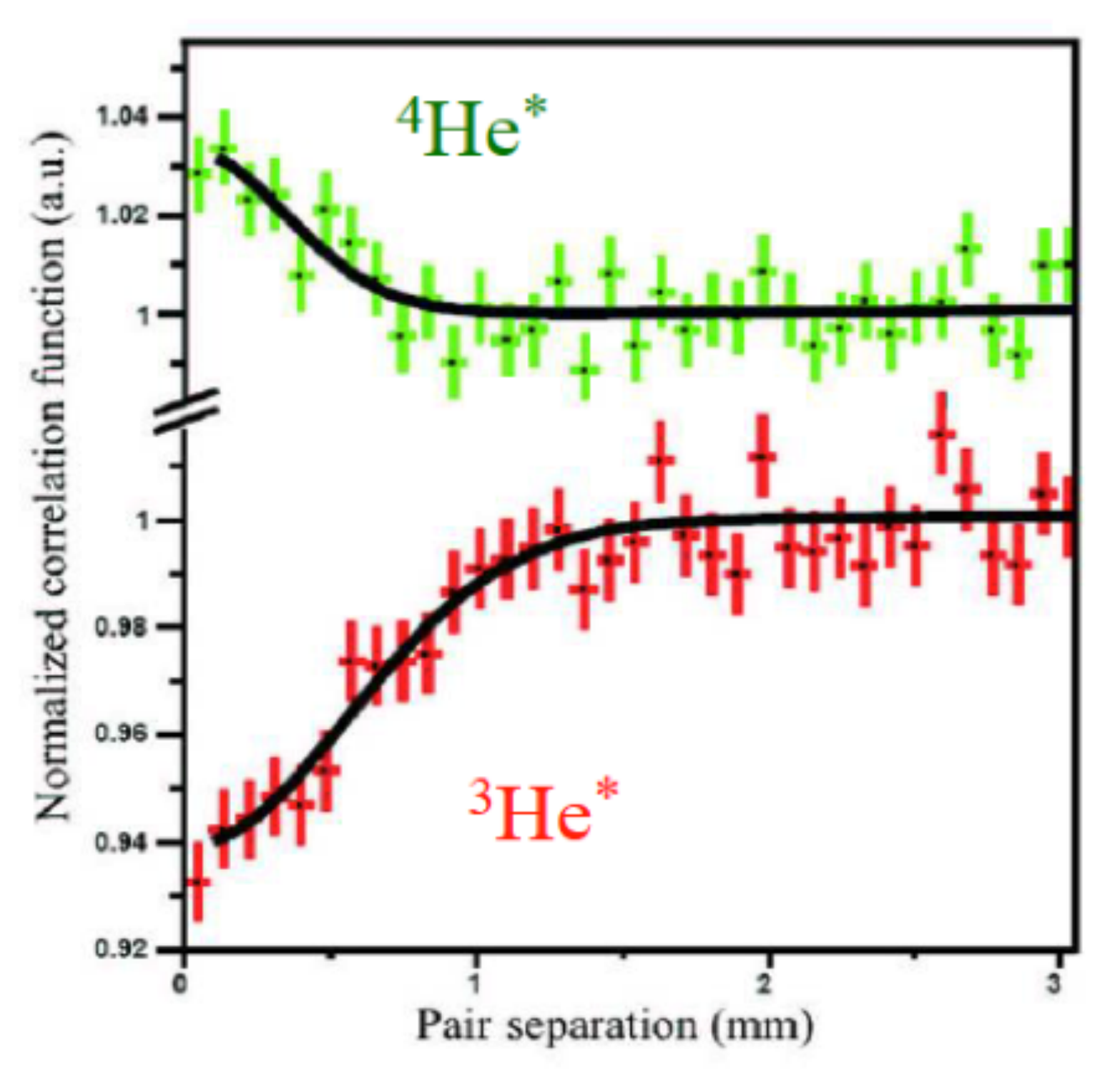}
\caption{Comparison of the HBT effect with bosons and fermions. Bosons tend to bunch while fermions tend to antibunch. The differences in width and amplitude of the dip vs. the bump are related to the difference in de Broglie wavelengths, which are in ratio 4 to 3 for atoms with the same velocity.}
\label{Figure_HBTfermion}
\end{figure}

While photons are bosons, atoms can come either in bosonic or in fermionic forms. The experiment described in section \ref{ss33} was performed with $^4$He atoms, which are bosons, but we have also performed a similar experiment with $^3$He atoms, which are fermions. The result is shown on Figure \ref{Figure_HBTfermion}, extracted from \cite{Jeltes:2007bu}.
 In that experiment, carried out in collaboration with colleagues at the VU of Amsterdam, we were able to realize a direct comparison of the effects for $^4$He and $^3$He atoms, initially held in the same trap at the same temperature, i.e., in clouds with identical shapes and widths. We could choose at will to drop either of the two isotopes. The density was small enough that interaction energy was negligible, and the observations result of quantum statistical effects only. One clearly sees that in the case of fermions one has a dip around zero rather than a bump. This is easily understood by referring to Figure \ref{Figure_toymodel}. In the case of fermions, the (entangled) state describing the two particles propagating from source to detectors must be antisymmetrized rather than symmetrized, so the amplitudes associated with the two diagrams of the right panel of Figure \ref{Figure_toymodel} must be added with opposite signs. For detectors  close enough to each other, and atoms with almost equal velocities so that they are undistinguishable,  it results into a null probability of joint detection, in agreement with the Pauli principle. Analogous results have been reported  in \cite{Rom:2006qs} for $^{40}$K atoms, and, with a  lower visibility,  for electrons in solids (see \cite{kiesel2002observation} and references in). 
 
In conclusion of that section, it must emphasized that there is absolutely no classical interpretation for the HBT effect with fermions. This is in contrast with the case of light, for which the  HBT  can be understood classically as a consequence of the Cauchy-Schwarz inequality for light intensity 
\begin{equation}
\left\langle {I^2} \right\rangle  \ge \left\langle I \right\rangle ^2 \,.
\label{eq11}
\end{equation}
In analogy, the HBT effect for fermions would be associated with the following inequality for atomic density
\begin{equation}
\left\langle {n{^2}} \right\rangle  < {\left\langle {n} \right\rangle ^2} \,.
\label{eq12}
\end{equation}
This is mathematically impossible if $n$ is a classical quantity, but it becomes possible in the framework of second quantization, where $n$ is considered an operator expressed as a function of the creation and annihilation operators for fermions.
\section{The Hong Ou and Mandel effect for photons}\label{s4}  
\subsection{\label{ss41} A two photon interference effect}

\begin{figure}[h!] 
\centering
\includegraphics[width=1.0\linewidth]{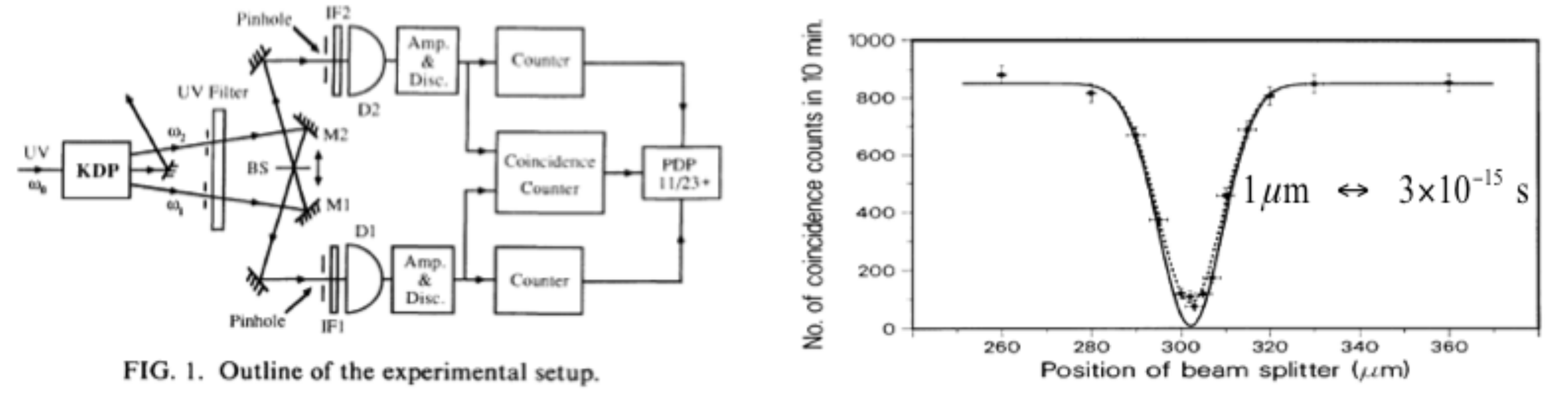}
\caption{Observation of the HOM dip in the joint detection of two photons emitted in  parametric down conversion and recombined on a beam-splitter. The probability of joint detection drops to zero when the two photon wave-packets arrive exactly at the same time on the beam-splitter. The width of the dip, of about 50~fs, indicates a simultaneity at that scale, a time resolution not previously heard of. }
\label{Figure_HOM}
\end{figure}

\begin{figure}[h!] 
\centering
\includegraphics[width=0.6\linewidth]{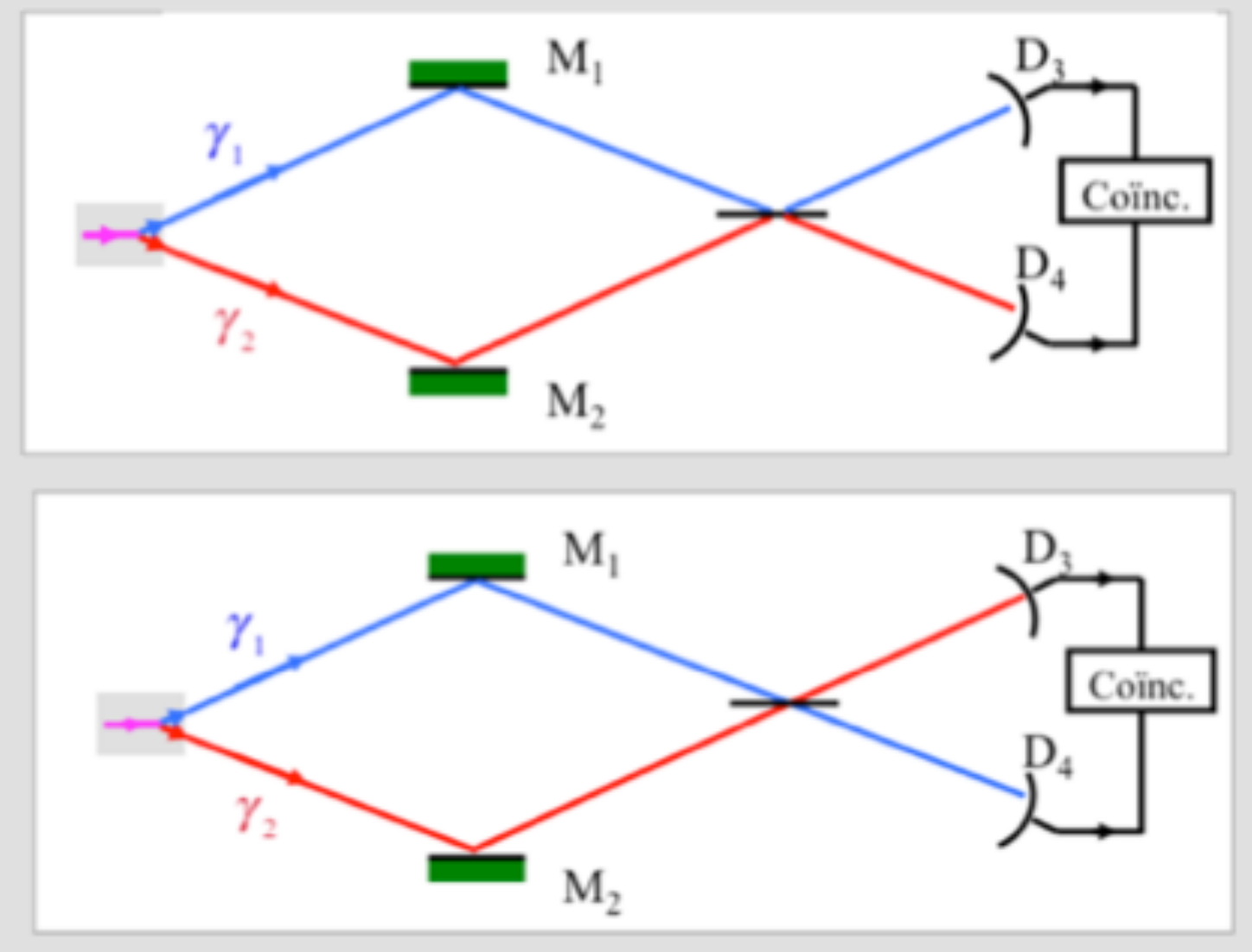}
\caption{The HOM effect: an emblematic two photon interference effect. If the two photons are undistinguishable and  exactly overlap at the balanced beam splitter, the two processes sketched on the two panels are undistinguishable, and their amplitudes must be added. It turns out that these amplitudes have the same modulus and opposite signs, so the interference is destructive, and the probability of a joint detection is null. One can equivalently understand the phenomenon by writing the state of the two photons as $\left| \Psi \right\rangle= 1/\sqrt{2}\, \left[ \left| 2_3,0_4 \right\rangle +  \left| 0_3,2_4 \right\rangle  \right] 
$, which is a (NOON) entangled state 
 ($\left| 2_3,0_4 \right\rangle$ means 2 photons in mode 3 and 0 photon in mode 4, propagating respectively from the beam splitter to detector D$_3$ or D$_4$, etc...).
}
\label{Figure_HOM_model}
\end{figure}

The Hong Ou and Mandel (HOM) effect was first described in a paper\cite{Hong:1987co} emphasizing its use to determine with a high resolution the ``simultaneity'' of the two photons (Figure \ref{Figure_HOM}) of pairs emitted in parametric down conversion from a CW laser beam, another landmark in quantum optics\cite{Burnham:1970dv}. Nowadays, the HOM effect is mostly cited as an emblematic example of a two photon interference effect, as shown on Figure \ref{Figure_HOM_model}. Indeed, 
a joint detection at D$_3$ and D$_4$ corresponds to two possible processes, which will interfere if the two photons are indistinguishable, i.e., if the two wave packets exactly overlap at the beam-splitter. A careful examination of the situation shows that for a balanced splitter the two two-photon amplitudes are opposite, so that the interference yields a null probability of joint detections. The opposite signs are related to  the unitarity of the matrix describing the effect of the beam-splitter. More precisely, if we choose the phase references such that all amplitude reflection and transmission  coefficients be real, the transmission coefficients involved in the lowest panel are equal, but the two  reflection coefficients involved in the upper panel have opposite signs. 

\subsection{\label{ss42} A fully quantum effect}
The HOM effect is the observation  that both photons are always detected in the same channel, either the upper one or the lower one, and never one in one channel and one in the other channel. This is an intriguing effect. If we thought of photons as classical particles with equal probabilities to be transmitted or reflected, the probability to observe a joint detection would be 1/2, while the probability of detecting both photons in the upper channel would be 1/4 and similarly for the probability to detect both photons in the lower channel. 

By analogy with the HBT effect, one might think that a semi-classical model involving classical waves could render an account of the situation. Let us indeed consider two classical waves with the same frequency entering in the two inputs of the balanced beam splitter. If their phase difference is $\phi$, their interference  leads to  rates  of single  detections in the output channels, $w^{(1)}(\mathrm{D}_3)$ and $w^{(1)}(\mathrm{D}_4)$, respectively proportional to $\sin^2 \phi$ and $\cos^2 \phi$. The rate of joint detections $w^{(2)}(\mathrm{D}_3;\mathrm{D}_4)$ is thus proportional to $\sin^2 \phi \cos^2 \phi=1/4 \sin^2 2\phi$. In order to render an account of the random character of the detections in either channel, we take $\phi$ a random variable uniformly distributed over an interval of $2\pi$. The average rates of single detection are then 1/2 each, while the average rate of joint detection is 1/8, i.e., $w^{(2)}(\mathrm{D}_3;\mathrm{D}_4)=1/2\,w^{(1)}(\mathrm{D}_3)\cdot w^{(1)}(\mathrm{D}_4)$. There is thus a suppression of the joint detection in $\mathrm{D}_3$ and $\mathrm{D}_4$, since the interference favors double detections in the same channel. That suppression, however,  is limited to a factor 1/2, while the quantum calculation leads to a total suppression, in agreement with the observation. A quantum  calculation of the shape of the dip obtained with parametric down conversion pairs can be found, for instance, in section 7.4.6 of \cite{grynberg2010introduction}.

It is remarkable that the effect can be generalized to the case where the two indistinguishable photon wave packets come from different sources. For instance, it has been observed with two spontaneous photons emitted by two different atoms, and terminating by chance in two modes of the electromagnetic field exact images of each other in the beam splitter\cite{Beugnon:2006od}. A calculation of the dip obtained with two independent one-photon wave packets can be found in Complement 5B of \cite{grynberg2010introduction}.
\section{The  Hong Ou and Mandel effect for atoms}\label{s5}  
In order to revisit, with atoms rather than photons,  quantum optics landmarks that are based on pairs of photons, we have developed a versatile source of pairs of $^4$He* atoms. This source is somewhat analogous to the sources of pairs of photons based on parametric down conversion of laser photons in a non-linear crystal\cite{Burnham:1970dv}. We start with a Bose Einstein Condensate of $^4$He*, dense enough that interaction energy between atoms plays a role analogous to a $\chi^{(3)}$ non-linearity for light. According to a suggestion of \cite{Molmer:2006ui}, first demonstrated in \cite{campbell2006parametric},  we apply a moving laser standing wave on a 1D interacting BEC in order  to favor the emission of pairs with well defined velocities\cite{bonneau2013tunable}. The phenomenon favoring this velocities selection is conservation of energy and quasi-momentum in the periodic potential provided by the standing wave. It is analogous to  phase matching in non-linear optics in non linear crystals. The non-linear process responsible for the emission of pairs is a dynamical instability associated with a repulsive interaction between atoms.
\begin{figure}[h!] 
\centering
\includegraphics[width=1.0\linewidth]{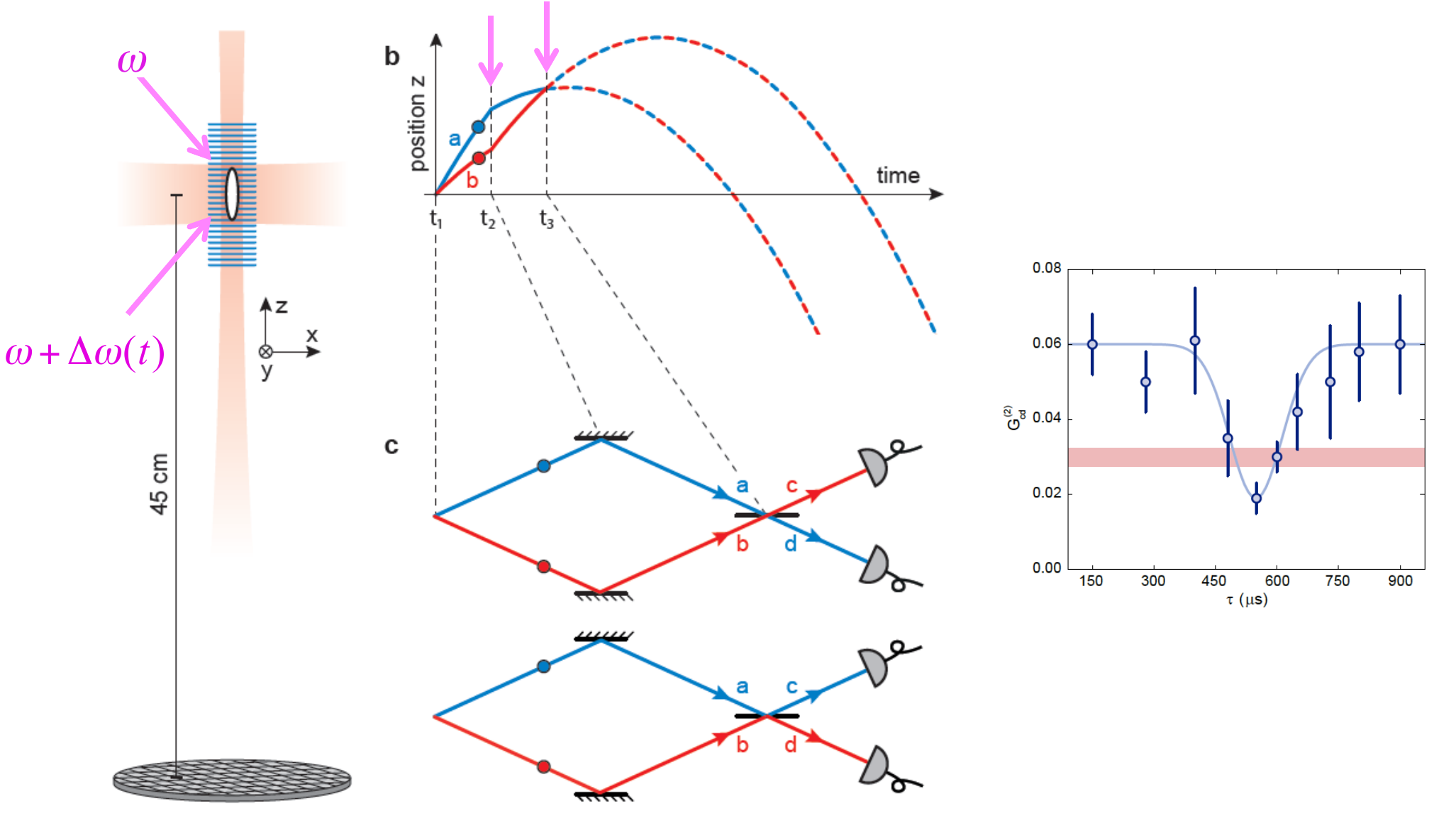}
\caption{Atomic HOM effect.}
\label{Figure_HOM_atom}
\end{figure}

With this source, we have implemented the experiment described on Figure \ref{Figure_HOM_atom}. The atoms are allowed to circulate along the vertical $z$ axis, and held on that axis by a far-off red detuned laser beam propagating along $z$. At time $t_0$, we apply the standing wave entailing the emission of  a pair of atoms  with controlled vertical velocities, as explained above. Their motion in gravity is represented as parabolas  in the $z,t$ diagram (panel b of Figure \ref{Figure_HOM_atom}). 

If now we consider the equivalent diagram  in a frame of reference falling freely as the center of mass of the two atoms, the motion is represented by two symmetrical strait lines (panel c of Figure \ref{Figure_HOM_atom}). At time $t_1$, we apply another laser standing wave, stationary in the free falling frame of reference thanks to the chirp $\Delta \omega(t)$ shown in the figure. That second standing wave realizes a Bragg diffraction of each atom, whose velocities are reverted if the standing wave is applied during a time and with an amplitude corresponding to a $\pi$ pulse. At time $t_2$ such that $t_2-t_1=t_1-t_0$, we apply again the second laser standing wave, still stationary in the free falling frame of reference, but for half the time only, realizing a $\pi/2$ pulse, which is equivalent to a balanced beamsplitter. 

As shown in the  panel c of Figure \ref{Figure_HOM_atom}, when viewed in the free falling frame of reference, the process is  equivalent to the one shown for photons in Figure \ref{Figure_HOM_model}, and if the atoms are in indistinguishable modes of matter-waves, we  observe the HOM dip. This is shown in the right panel of Figure \ref{Figure_HOM_atom}, where the time delay is controlled by changing the time $t_2$ around the value given by $t_2-t_1=t_1-t_0$. 

Note that in the experiment with atoms, we have a bunch of atoms submitted to the process, and we do not have two detectors, but we have the equivalent of  many detectors monitoring all the atoms (cf. subsection \ref{ss32} 	and Figure
\ref{Figure_He}). Among all the registered detections, we then look a posteriori for pairs of atoms corresponding to modes symmetrical in the final beam splitter. Compared to the case of photons, where corresponding modes are selected a priori with pinholes placed before the mirrors and beamsplitter, our selection is done a posteriori, which is possible thanks to our many pixels detectors.

The fact that the dip does not go to zero is fully accounted for by the fact that in the pair creation process there is some amplitude to have 2 atoms rather than 1 in an elementary mode. The amount of that contamination can be determined by using the stored data to calculate the $g^{(2)}$ function in an elementary mode. For one atom only, that function should be zero, but we find it different from zero and infer the amplitude for 2 atoms in the mode. Even with that imperfection of our experiment, the visibility of the dip is larger than 1/2, which means that the observed effect could not be explained by ``ordinary'' interferences of atomic matter-waves in the ordinary space-time, and demands an interpretation in terms of two atom amplitudes.  
\section{Outlook: towards Bell's inequalities test with atoms}\label{s6}  
\begin{figure}[h!] 
\centering
\includegraphics[width=0.9\linewidth]{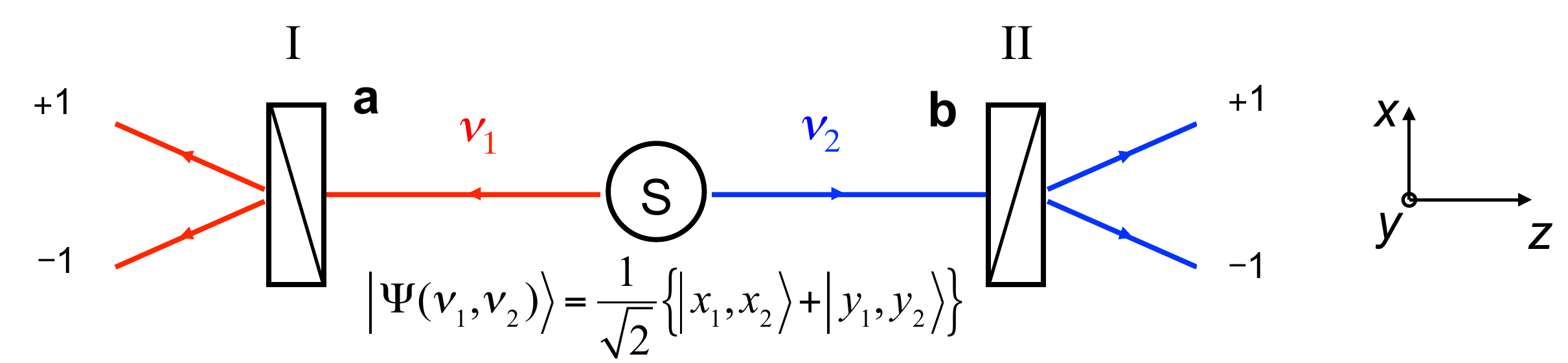}
\caption{Bell's inequalities test with well separated photons. Having two modes ($\left|x_i \right\rangle$ and $\left|y_i \right\rangle$) for each separated photon ($i=1$ or $i=2$)  allows one to choose at will,  for each photon,  between different directions of polarization measurement. If the polarizers are far enough from each other, and can be adjusted fast enough, the choice can be made while the photons are in flight, enforcing  loclaity, i.e., relativistic separation between the measurements. A test of Bell's inequalities in such a configuration allows one to decide  between Einstein's Local Realism and Quantum Mechanics.  }
\label{Figure_Bell}
\end{figure}

Experimenting on light has played a major role in the development of both the first and second quantum revolution. Wave particle duality for light was recognized by Einstein as early as 1909\cite{Einstein:1909oa}, while it was only in 1923 that Louis de Broglie proposed that wave-particle duality should also apply to material particles. When it comes to the second quantum revolution, of which   entanglement and two particles interference effects are  key ingredients,  light is again far ahead, with first violations of   Bell's inequalities reported in the early 1970's, and the  conflict with relativistic locality  demonstrated in the early 1980's \cite{Aspect:1999gm}, with polarizers varied during the flight of photons (Figure \ref{Figure_Bell}). A new generation of experiments,  started in the late 1990's, has lead to improved tests, culminating in 2015 with 
almost perfect, so-called loophole-free, experiments\cite{aspect2015viewpoint}.  In contrast, no Bell's  inequalities tests have been performed on the external degrees of freedom (position or momentum) of material particles\footnote{Tests of Bell's inequalities with ions of \cite{rowe2001experimental} bore on internal degrees of freedom.}.  
Such tests would be highly desirable, not only because they would complete the long series of  correspondance between landmarks of photon optics and of atom optics [Table \ref{Table_photon_atom}], but also because they may open the way to experiments that could shed some light on the elusive frontier between quantum physics and relativity. 

One may wonder why the HOM scheme is not sufficient to carry out such investigations. 
The reason is that the HOM effect does not address the question of non-locality, i.e., the tension between relativity and quantum mechanics, which was a major element in the EPR argument against the completeness of Quantum Mechanics. This is because  two modes only (3 and 4) are involved  in the entangled state of the two HOM particles
\begin{equation}
\left| \Psi \right\rangle  = \frac{1}{\sqrt 2} \left[ \left| 2_3, 0_4 \right\rangle + \left| 0_3, 2_4 \right\rangle     \right]
 \,.
\label{eq13}
\end{equation}
while  Bell's inequalities tests demand to have an entangled state of two particles in four modes, such as
\begin{equation}
\left| \Psi \right\rangle  = \frac{1}{\sqrt 2} \left[ \left| x_1,x_2 \right\rangle + \left| y_1,_2 \right\rangle     \right]
 \,,
\label{eq14}
\end{equation}
(see Figure \ref{Figure_Bell}). More precisely in order to test locality, one must be able to choose between two non-commuting observables for each of  two spatially separated particles, and this demands a two-dimensional space on each side, for each particle.\footnote{One can note that in contrast to Bell's inequalities tests,  the quantum behavior in the HOM experiment can be mimicked by a local hidden variable theory where the two photons are determined, from the moment of the emission, to both go either on one side or the other side.}  

\begin{figure}[h!] 
\centering
\includegraphics[width=0.8\linewidth]{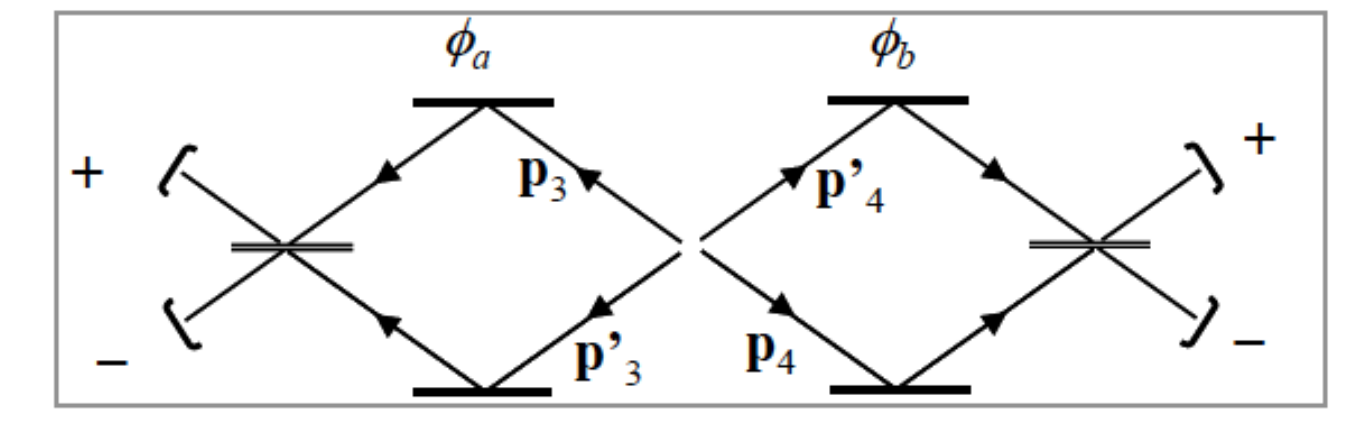}
\caption{Proposed configuration to test Bell's inequalities with two atoms in the momentum entangled  state $\left| \Psi \right\rangle  = \frac{1}{\sqrt 2} \left[ \left| p_3,p_4\right\rangle + \left| p'_3,p'_4 \right\rangle     \right]$  }
\label{Figure_rarity}
\end{figure}

As a fist step towards a Bell's test with massive particles entangled in momentum,  the experiment described in section \ref{s5} has allowed us to find an evidence of entanglement of two atoms in 4 modes associated with different momenta\cite{dussarrat2017two}. This is hopefully the last step towards a genuine test of Bell's inequalities with a pair of material particles entangled in momentum, following a scheme in the spirit of the experiment of \cite{Rarity:1990mh}, sketched in Figure \ref{Figure_rarity}. It  would complete the series of landmark quantum optics experiments revisited with atoms, and start a series of such experiments with heavier particles, allowing one  to address the interface between quantum physics and relativity. 

\begin{table}
\centering
\begin{tabular}{|c|c||c|c|}
\hline
{\bf Photon Optics} & date & {\bf Atom Optics} & date\\ \hline\hline
Interference, diffraction			& 1800's 		&Interference, diffraction		& 1990's \\
Single photons 					& 1974,1985	&Single atoms				& 2002\\
Photon correlations (HBT)			&1955		&Atom correlations (HBT)		&2005\\
$\chi^{(2)}$ photon pairs			&1970's		&$\chi^{(3)}$ atom pairs		&2007\\
Beyond SQL: squeezing			& 1985 		&Beyond SQL: squeezing		&2010 \\
Bell tests spontaneous  photons 	& 1972,1982	&Bell tests molecules dissoc.	&?\\
HOM with 	$\chi^{(2)}$ pairs		&1987		&HOM with $\chi^{(2)}$ pairs	&2014\\
Bell tests with $\chi^{(2)}$ pairs		&1989-98		&Bell tests with $\chi^{(3)}$ pairs	&?\\
\hline\hline
\end{tabular}
\caption{\textbf{Photon vs. Atom quantum optics: some landmarks.} 
``SQL'': Standard Quantum Limits. ``?'': not (yet) done.}\label{Table_photon_atom}
\end{table}

\bibliographystyle{ieeetr}
\bibliography{/Applications/TeX/bibliotheques.bib/biblioAA_6}

\begin{thebibliography}{10}

\bibitem{lamb1964theory}
W.~E. Lamb~Jr, ``Theory of an optical maser,'' {\em Physical Review}, vol.~134,
  no.~6A, p.~A1429, 1964.

\bibitem{grynberg2010introduction}
G.~Grynberg, A.~Aspect, and C.~Fabre, {\em Introduction to quantum optics: from
  the semi-classical approach to quantized light}.
\newblock Cambridge university press, 2010.

\bibitem{Aspect:1999gm}
A.~Aspect, ``Bell's inequality test: more ideal than ever,'' {\em Nature},
  vol.~398, no.~6724, pp.~189--190, 1999.

\bibitem{aspect2015viewpoint}
A.~Aspect, ``Viewpoint: Closing the door on einstein and bohr's quantum
  debate,'' {\em Physics}, vol.~8, p.~123, 2015.

\bibitem{Hong:1987co}
C.~K. Hong, Z.~Y. Ou, and L.~Mandel, ``Measurement of subpicosecond time
  intervals between 2 photons by interference,'' {\em Physical Review Letters},
  vol.~59, no.~18, pp.~2044--2046, 1987.

\bibitem{aspect:2017mooc1photon}
A.~Aspect, ``Mooc: Quantum optics 1 : single photons,'' {\em
  https://www.coursera.org/learn/quantum-optics-single-photon}, 2017.

\bibitem{Kimble:1977kk}
H.~J. Kimble, M.~Dagenais, and L.~Mandel, ``Photon anti-bunching in resonance
  fluorescence,'' {\em Physical Review Letters}, vol.~39, no.~11, pp.~691--695,
  1977.

\bibitem{Grangier:1986ie}
P.~Grangier, G.~Roger, and A.~Aspect, ``Experimental-evidence for a photon
  anticorrelation effect on a beam splitter - a new light on single-photon
  interferences,'' {\em Europhysics Letters}, vol.~1, no.~4, pp.~173--179,
  1986.

\bibitem{dowling2003quantum}
J.~P. Dowling and G.~J. Milburn, ``Quantum technology: the second quantum
  revolution,'' {\em Philosophical Transactions of the Royal Society of London
  A: Mathematical, Physical and Engineering Sciences}, vol.~361, no.~1809,
  pp.~1655--1674, 2003.

\bibitem{Aspect:2004introductionsuqm}
A.~Aspect, {\em Introduction: John Bell and the second quantum revolution},
  pp.~xvii--xl.
\newblock Cambridge: Cambridge University Press, 2~ed., 2004.

\bibitem{Feynman:1963kf}
R.~P. Feynman, {\em Lectures on Physics}.
\newblock Addison-Wesley, 1963.

\bibitem{Feynman:1982yx}
R.~P. Feynman, ``Simulating physics with computers,'' {\em International
  Journal of Theoretical Physics}, vol.~21, no.~6-7, pp.~467--488, 1982.

\bibitem{peres1995quantum}
A.~Peres, {\em Quantum theory: concepts and methods}, vol.~57.
\newblock Springer, 1995.

\bibitem{brown1956correlation}
R.~H. Brown and R.~Twiss, ``Correlation between photons in two coherent beams
  of light,'' {\em Nature}, vol.~177, no.~4497, pp.~27--29, 1956.

\bibitem{forrester1955photoelectric}
A.~T. Forrester, R.~A. Gudmundsen, and P.~O. Johnson, ``Photoelectric mixing of
  incoherent light,'' {\em Physical Review}, vol.~99, no.~6, p.~1691, 1955.

\bibitem{brown1974intensity}
R.~H. Brown, ``The intensity interferometer: its application to astronomy,''
  {\em Research supported by the Department of Scientific and Industrial
  Research, Australian Research Grants Committee, US Air Force, et al. London,
  Taylor and Francis, Ltd.; New York, Halsted Press, 1974. 199 p.}, 1974.

\bibitem{Rebka:1957oy}
G.~A. Rebka and R.~V. Pound, ``Time-correlated photons,'' {\em Nature},
  vol.~180, no.~4594, pp.~1035--1036, 1957.

\bibitem{glauber1963photon}
R.~J. Glauber, ``Photon correlations,'' {\em Physical Review Letters}, vol.~10,
  no.~3, pp.~84--86, 1963.

\bibitem{glauber1963coherent}
R.~J. Glauber, ``Coherent and incoherent states of the radiation field,'' {\em
  Physical Review}, vol.~131, no.~6, p.~2766, 1963.

\bibitem{glauber1963quantum}
R.~J. Glauber, ``The quantum theory of optical coherence,'' {\em Physical
  Review}, vol.~130, no.~6, p.~2529, 1963.

\bibitem{glauber1965houches}
R.~Glauber, ``Les houches lecture notes, 1964 (quantum optics and electronics,
  ed. c. de witt et al., gordon and breach, ny, 1965).'2) g,'' {\em Lachs:
  Phys. Rev}, vol.~138, p.~B1012, 1965.

\bibitem{purcell1994reproduced}
E.~Purcell, ``Reproduced fromnature (1956) 17 8, 1449--50,'' {\em Journal of
  Astrophysics and Astronomy}, vol.~15, no.~1, pp.~27--32, 1994.

\bibitem{baym1998physics}
G.~Baym, ``The physics of hanbury brown--twiss intensity interferometry: from
  stars to nuclear collisions,'' {\em arXiv preprint nucl-th/9804026}, 1998.

\bibitem{Yasuda:1996po}
M.~Yasuda and F.~Shimizu, ``Observation of two-atom correlation of an ultracold
  neon atomic beam,'' {\em Physical Review Letters}, vol.~77, no.~15,
  pp.~3090--3093, 1996.

\bibitem{Schellekens:2005ql}
M.~Schellekens, R.~Hoppeler, A.~Perrin, J.~V. Gomes, D.~Boiron, A.~Aspect, and
  C.~I. Westbrook, ``Hanbury brown twiss effect for ultracold quantum gases,''
  {\em Science}, vol.~310, no.~5748, pp.~648--651, 2005.

\bibitem{arecchi1966time}
F.~Arecchi, E.~Gatti, and A.~Sona, ``Time distribution of photons from coherent
  and gaussian sources,'' {\em Physics Letters}, vol.~20, no.~1, pp.~27--29,
  1966.

\bibitem{Ottl:2005ai}
A.~Ottl, S.~Ritter, M.~Kohl, and T.~Esslinger, ``Correlations and counting
  statistics of an atom laser,'' {\em Physical Review Letters}, vol.~95, no.~9,
  2005.

\bibitem{hodgman2011direct}
S.~Hodgman, R.~Dall, A.~Manning, K.~Baldwin, and A.~Truscott, ``Direct
  measurement of long-range third-order coherence in bose-einstein
  condensates,'' {\em Science}, vol.~331, no.~6020, pp.~1046--1049, 2011.

\bibitem{Altman:2004ac}
E.~Altman, E.~Demler, and M.~D. Lukin, ``Probing many-body states of ultracold
  atoms via noise correlations,'' {\em Physical Review A}, vol.~70, no.~1,
  2004.

\bibitem{Jeltes:2007bu}
T.~Jeltes, J.~M. McNamara, W.~Hogervorst, W.~Vassen, V.~Krachmalnicoff,
  M.~Schellekens, A.~Perrin, H.~Chang, D.~Boiron, A.~Aspect, and C.~I.
  Westbrook, ``Comparison of the hanbury brown-twiss effect for bosons and
  fermions,'' {\em Nature}, vol.~445, no.~7126, pp.~402--405, 2007.

\bibitem{Rom:2006qs}
T.~Rom, T.~Best, D.~van Oosten, U.~Schneider, S.~Folling, B.~Paredes, and
  I.~Bloch, ``Free fermion antibunching in a degenerate atomic fermi gas
  released from an optical lattice,'' {\em Nature}, vol.~444, no.~7120,
  pp.~733--736, 2006.

\bibitem{kiesel2002observation}
H.~Kiesel, A.~Renz, and F.~Hasselbach, ``Observation of hanbury brown--twiss
  anticorrelations for free electrons,'' {\em Nature}, vol.~418, no.~6896,
  pp.~392--394, 2002.

\bibitem{Burnham:1970dv}
D.~C. Burnham and D.~L. Weinberg, ``Observation of simultaneity in parametric
  production of optical photon pairs,'' {\em Physical Review Letters}, vol.~25,
  no.~2, pp.~84--87, 1970.

\bibitem{Beugnon:2006od}
J.~Beugnon, M.~P.~A. Jones, J.~Dingjan, B.~Darquie, G.~Messin, A.~Browaeys, and
  P.~Grangier, ``Quantum interference between two single photons emitted by
  independently trapped atoms,'' {\em Nature}, vol.~440, no.~7085,
  pp.~779--782, 2006.

\bibitem{Molmer:2006ui}
K.~Molmer, ``Phase-matched matter wave collisions in periodic potentials,''
  {\em New Journal of Physics}, vol.~8, 2006.

\bibitem{campbell2006parametric}
G.~K. Campbell, J.~Mun, M.~Boyd, E.~W. Streed, W.~Ketterle, and D.~E.
  Pritchard, ``Parametric amplification of scattered atom pairs,'' {\em
  Physical review letters}, vol.~96, no.~2, p.~020406, 2006.

\bibitem{bonneau2013tunable}
M.~Bonneau, J.~Ruaudel, R.~Lopes, J.-C. Jaskula, A.~Aspect, D.~Boiron, and
  C.~I. Westbrook, ``Tunable source of correlated atom beams,'' {\em Physical
  Review A}, vol.~87, no.~6, p.~061603, 2013.

\bibitem{Einstein:1909oa}
A.~Einstein, ``On the evolution of our vision on the nature and constitution of
  radiation,'' {\em Physikalische Zeitschrift}, vol.~10, pp.~817--826, 1909.

\bibitem{rowe2001experimental}
M.~A. Rowe, D.~Kielpinski, V.~Meyer, C.~A. Sackett, W.~M. Itano, C.~Monroe, and
  D.~J. Wineland, ``Experimental violation of a bell's inequality with
  efficient detection,'' {\em Nature}, vol.~409, no.~6822, pp.~791--794, 2001.

\bibitem{dussarrat2017two}
P.~Dussarrat, M.~Perrier, A.~Imanaliev, R.~Lopes, A.~Aspect, M.~Cheneau,
  D.~Boiron, and C.~I. Westbrook, ``Two-particle four-mode interferometer for
  atoms,'' {\em Physical Review Letters}, vol.~119, no.~17, p.~173202, 2017.

\bibitem{Rarity:1990mh}
J.~G. Rarity and P.~R. Tapster, ``Experimental violation of bell inequality
  based on phase and momentum,'' {\em Physical Review Letters}, vol.~64,
  no.~21, pp.~2495--2498, 1990.

\end{thebibliography}
%

%
%
%

\end{document}